\documentclass{jpsj-suppl}
\usepackage{times}
\usepackage{color}
\usepackage{graphicx}
\usepackage{amsmath}
\usepackage{amssymb}
\usepackage{ulem}

\def\e{\epsilon}

\def\S{\Sigma}
\def\s{\sigma}

\def\w{\omega}
\def\ua{\uparrow}
\def\da{\downarrow}

\def\Vec#1{\mathbf #1}

\title{Doping evolution of the electron-hole asymmetric $s$-wave pseudogap in underdoped high-$T_c$ cuprate superconductors}

\author{Shiro \textsc{Sakai}$^{1}$ and Marcello \textsc{Civelli}$^{2}$}

\inst{$^{1}$Department of Applied Physics, University of Tokyo, 7-3-1 Hongo, Bunkyo-ku, Tokyo 113-8656, Japan\\
$^{2}$Laboratoire de Physique des Solides, Universit\'e Paris-Sud, CNRS, UMR 8502, F-91405 Orsay Cedex, France}

\abst{We study the doping evolution of the electronic structure in the pseudogap state of high-$T_c$ cuprate superconductors, by means of a cluster extension of the dynamical mean-field theory applied to the two-dimensional Hubbard model.
The calculated single-particle excitation spectra in the strongly underdoped regime show a marked electron-hole asymmetry and reveal a ``$s$-wave'' pseudogap, which display a finite amplitude in all the directions in the momentum space but not always at the Fermi level: The energy location of the gap strongly depends on momentum, and in particular in the nodal region, it is {\it above} the Fermi level.
With increasing hole doping, the pseudogap disappears everywhere in the momentum space.
We show that the origin and the ``{\it s}-wave'' structure of the pseudogap  can be ascribed to the emergence of a strong-scattering surface, which appears in the energy-momentum space close to the Mott insulator.}

\kword{cuprate, superconductivity, pseudogap, electron-hole asymmetry, dynamical mean-field theory}

\begin{document}
\maketitle

\section{Introduction}

The high-$T_c$ cuprate superconductors show many anomalous behaviors already in the metallic state above the critical temperature $T_c$.
Various spectroscopy experiments have elucidated that the electronic structure in the anomalous metallic state indeed deviates from that of standard metals in many respects.
The deviation is especially conspicuous in underdoped samples, for which angle-resolved photo-emission spectroscopy (ARPES)  \cite{damascelli03} has observed a momentum-dependent gap (pseudogap) up to a temperature $T^\ast$ significantly higher than $T_c$ \cite{loeser96,ding96,norman98}.
The origin of the pseudogap and its relationship to the superconductivity have been a central issue in the study of the cuprate superconductors \cite{norman05}, and a number of phenomenological theories have been proposed for explaining both the pseudogap and the superconductivity.
A prevailing assumption in these theories is that the pseudogap has a $d$-wave symmetry \cite{anderson87,emery95,chakravarty01,balents98,franz01,yang06}, in close analogy with the superconducting gap, which displays a $d$-wave structure, as established by quantum interference experiments \cite{wollman93,tsuei97}. This assumption finds comforting support in ARPES results, which find a well-developed pseudogap around the antinodal direction (the direction along Cu-O bonds), and the pseudogap closing in the nodal direction (45$^\circ$ from Cu-O bonds).
In our view, however, these ARPES results do not necesserily imply a $d$-wave symmetry of the pseudogap, because the gap structure in the unoccupied side of the spectra is virtually unrevealed by the ARPES. 
As a matter of facts, there are increasing evidences of an electron-hole asymmetry (a larger gap above the Fermi level than below it) in the pseudogap structure \cite{yang08,hashimoto10,he11}, which makes the conventional ``symmetrization" procedure \cite{norman98} of the ARPES spectra inappropriate for describing the electronic structure above the Fermi level.

\begin{figure}[t]
\begin{center}
\includegraphics[width=0.8\textwidth]{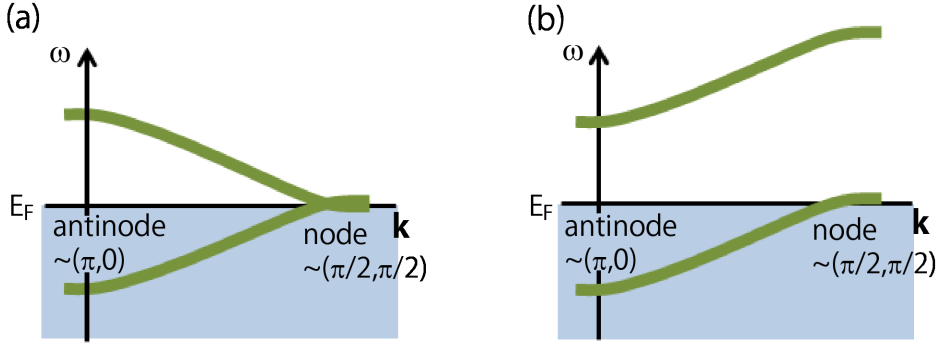}
\end{center}
\caption{(Color online) Schematic illustration of the difference between (a) the conventionally assumed $d$-wave and (b) the recently proposed $s$-wave pseudogap structures. The green curves represent the electronic bands and the pseudogap is the energy difference between them. The region below the Fermi level, $E_\text{F}$, is tinted by light blue. Notice that both structures look the same below $E_F$, where the ARPES applies, while they differ above $E_F$.
}
\label{f1}
\end{figure}

A pseudogap structure reconciling the $d$-wave-like structure below the Fermi level and the observed electron-hole asymmetry has been recently proposed in Ref.\cite{sakai10,alain13,sakai13}, on the basis of microscopic numerical simulations on the Hubbard model and its comparison with various experimental results.
Figure 1 schematically illustrates the two different points of view: the more conventional $d$-wave pseudogap [Fig.~1(a)] and the recently proposed ``$s$-wave" one [Fig.~1(b)].
Both share a similar $d$-wave-like structure {\it below} the Fermi level ($E_F$), and are consistent with ARPES observation of finite gap at the antinodes and zero gap at the nodes. However, the unoccupied side is different:
The $d$-wave pseudogap is symmetric with respect to the Fermi level and therefore the gap closes around the node, while the ``$s$-wave" pseudogap shows a strong electron-hole asymmetry and in particular the gap {\it above} $E_F$ does not close at the node. The structure presented in Fig.~1(b) is called (with some abuse of language) ``$s$-wave" because the gap amplitude is finite everywhere (from the antinode to the node) in the energy-momentum space.
This is distinct from the standard $s$-wave-gap definition, which implies a gap always present at the Fermi level.
 
The $s$-wave pseudogap is indeed supported by various numerical studies on the two-dimensional Hubbard model \cite{kyung06,sakai09,sakai10,sakai13} as well as on the $t$-$J$ model \cite{tohyama04}, which, {\it without making any assumption on the structure or symmetry (and even the presence) of the pseudogap,} show the $s$-wave structure of the pseudogap. 
The $s$-wave-pseudogap proposal also explains various spectral anomalies observed in experiments.
In particular, in strongly underdoped regime, it reproduces the electron-hole spectral asymmetry observed by ARPES analyses on the thermally populated states \cite{yang08} and on the shift of the back-bending bands \cite{hashimoto10,he11}, as well as the electron-hole asymmetry observed by the scanning tunneling microscopy (STM) \cite{hanaguri04,anderson06}. 
Moreover, we have shown\cite{sakai13} that electronic Raman response calculated within the framework of the $s$-wave pseudogap state are able to reproduce peculiar temperature dependences of the experimental Raman response. Since the Raman responses detect both the occupied and unoccupied parts of the electronic structure, and (by employing different light polarization) they can also access independently both the nodal and anti-nodal regions of momentuum-space, the agreement strongly supports the $s$-wave pseudogap.

In this paper we study the doping evolution of the electronic structure in the two-dimensional Hubbard model by means of the cellular dynamical mean-field theory (CDMFT) \cite{kotliar01}, and clarify how the $s$-wave pseudogap state present at small dopings evolves into the normal Fermi liquid state at high dopings. 
We also calculate the real-frequency self-energy, and show that the $s$-wave pseudogap and its anomalous energy-momentum structure can be attributed to the emergence of a strong scattering surface in the vicinity of the Mott insulator.

\section{Model and Method}

We study the two-dimensional Hubbard model,
\begin{align}
H= \sum_{\Vec{k}\s}\e(\Vec{k})c_{\Vec{k}\s}^\dagger c_{\Vec{k}\s}
-\mu \sum_{i\s}n_{i\s}+ U\sum_{i}n_{i\ua}n_{i\da},
\label{eq:hubbard}
\end{align}
on a square lattice.
Here $c_{\Vec{k}\s}$ $(c_{\Vec{k}\s}^\dagger)$ annihilates (creates) 
an electron of spin $\s$ with momentum $\Vec{k} = (k_x, k_y)$, 
$c_{i\s}$ $(c_{i\s}^\dagger)$ is its Fourier component at site $i$, and
$n_{i\s}\equiv c_{i\s}^\dagger c_{i\s}$.
$U$ represents the onsite Coulomb repulsion, $\mu$ the chemical
potential, and 
$\e(\Vec{k})\equiv -2t(\cos k_x + \cos k_y) -4t' \cos k_x\cos k_y,$
where $t$ $(t')$ is the (next-)nearest-neighbor transfer integral. 
The hole doping is defined by $1-n$, where $n\equiv \sum_{\s}\langle n_{i\s}\rangle$ is the electron density.
We adopt $t=0.3$eV, $t'=-0.2t$ and $U=8t$, which are reasonable values for hole-doped cuprates and indeed reproduce the Mott insulating state for the undoped case ($n=1$).

The CDMFT maps the model (\ref{eq:hubbard}) onto an effective Anderson model consisting of a cluster of $N_\text{C}$ interacting sites, embedded in an infinite bath of free electrons which describes a dynamical mean field\cite{kotliar01}. The effective cluster model is solved by the continuous-time quantum Monte Carlo method\cite{gull11}. In this work we employ an $N_\text{C}$$=$2$\times$2 square cluster since larger clusters suffer a severe sign problem, especially for dopings larger than 5\%.
The momentum-dependent quantities have been extracted from the cluster ones by using the cumulant interpolation scheme\cite{stanescu06,stanescu06-2} which restores the lattice periodicity broken by the lattice-partition into clusters. 
The cumulant periodization applied to 2$\times$2 cluster was previously shown to give a rather reasonable result, close to that of a larger 4$\times$3 or 4$\times$4 clusters, in a relevant parameter region \cite{sakai12}.
We direct readers to Ref.~\cite{sakai12} for further details on the methodology.

\section{Results and Discussion}

\begin{figure}
\begin{center}
\includegraphics[width=\textwidth]{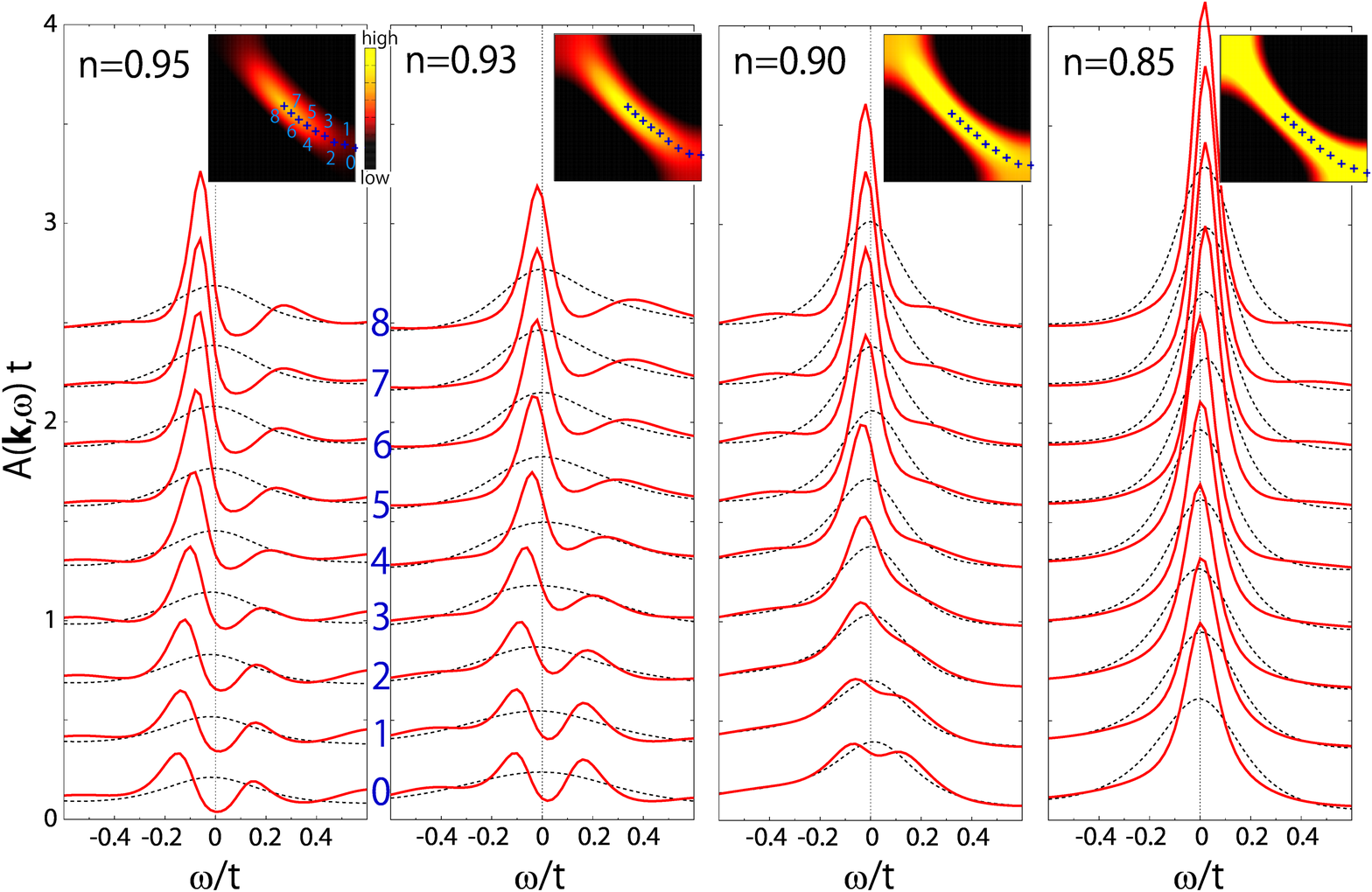}
\label{f2}
\end{center}
\caption{(Color online) The evolution of the one-particle excitation spectra from the antinode to the node.
Black dashed curves show the results at $T>T^\ast$; $T=0.15t$ for $n=0.95$ and $T=0.10t$ otherwise.
Red solid curves show the results at $T=0.04t < T^\ast$.
The curves are offset by 0.3 for clarity.
Inset: The momentum map of the low-energy spectral weight at $T=0.04t$. Blue crosses show the momenta where the spectral function is calculated.}
\end{figure}

Figure 2 shows the single-particle excitation spectra $A(\Vec{k},\w)$ calculated for various dopings at several momenta (marked by blue crosses in the inset of each panel), from the antinodal region to the nodal region.
These $\Vec{k}$-points have been determined by picking up the ones at which $A(\Vec{k},\w)$ (dashed black curves) shows a peak at the Fermi level ($\w=0$) at a temperature above $T^\ast$ (the pseudogap crossover temperature). This procedure follows in spirit the experimental approach often used in ARPES studies \cite{norman98,hashimoto10,he11}, where the underlying Fermi surface is determined by the $\Vec{k}$-points displaying the maximal spectral intensity (peaks) at $T > T^\ast$. 
Note that this does not necessarily correspond to the actual Fermi surface at zero temperature since the self-energy changes dramatically by lowering temperatures (see the following discussion on Fig. 2). The large self-energy at $T=0$ can transform the Fermi surface into hole pockets around the nodal points, as found for instance in the CDMFT + exact diagonalization studies \cite{stanescu06,stanescu06-2,civelli09,sakai09,sakai10}.
The goal of the present study is therefore to compare our CDMFT spectra on the Hubbard model with ARPES results on cuprates at moderately high temperatures. We leave open the questions about the evolution of the electronic structure at low temperatures, where other competing orders could come into play and determine different ground states. For example, quantum oscillations experiments on Y-Ba-Cu-O\cite{leboeuf07} and recent transport measures on a Hg-based cuprate \cite{doiron2013} upon application of strong magnetic fields, seem to indicate a charge-ordered state emerging in the underdoped region of the phase diagram.

At a high doping ($n=0.85$) the broad peak at a high temperature $T=0.10t > T^\ast$ (dashed black curves) evolves into a long-lived quasiparticle peak as the temperature is lowered down to $T=0.04t$ (solid red curves).
On the other hand, at lower dopings ($n=0.90, 0.93, 0.95$), there appears a gap (pseudogap) at a low energy, which becomes more pronounced as the doping is reduced. The gap is nearly centered around $\w=0$ and close to the antinode (momentum point 0), while it shifts to positive energy towards the node (point 8). This structure is compatible with Fig.~1(b), where we have introduced the "$s$-wave" electron-hole asymmetric pseudogap.
The strong electron-hole asymmetry which develops in between the nodal and antinodal points and the more particle-hole symmetric sprectra at the antinode are fairly consistent with the thermally-populated-state analysis of the ARPES study of Ref.\cite{yang08}. 
We also note that the pseudogap not only decreases in amplitude but also shows to be filled as the doping is increased.\cite{lin10,sordi12}

\begin{figure}
\begin{center}
\includegraphics[width=\textwidth]{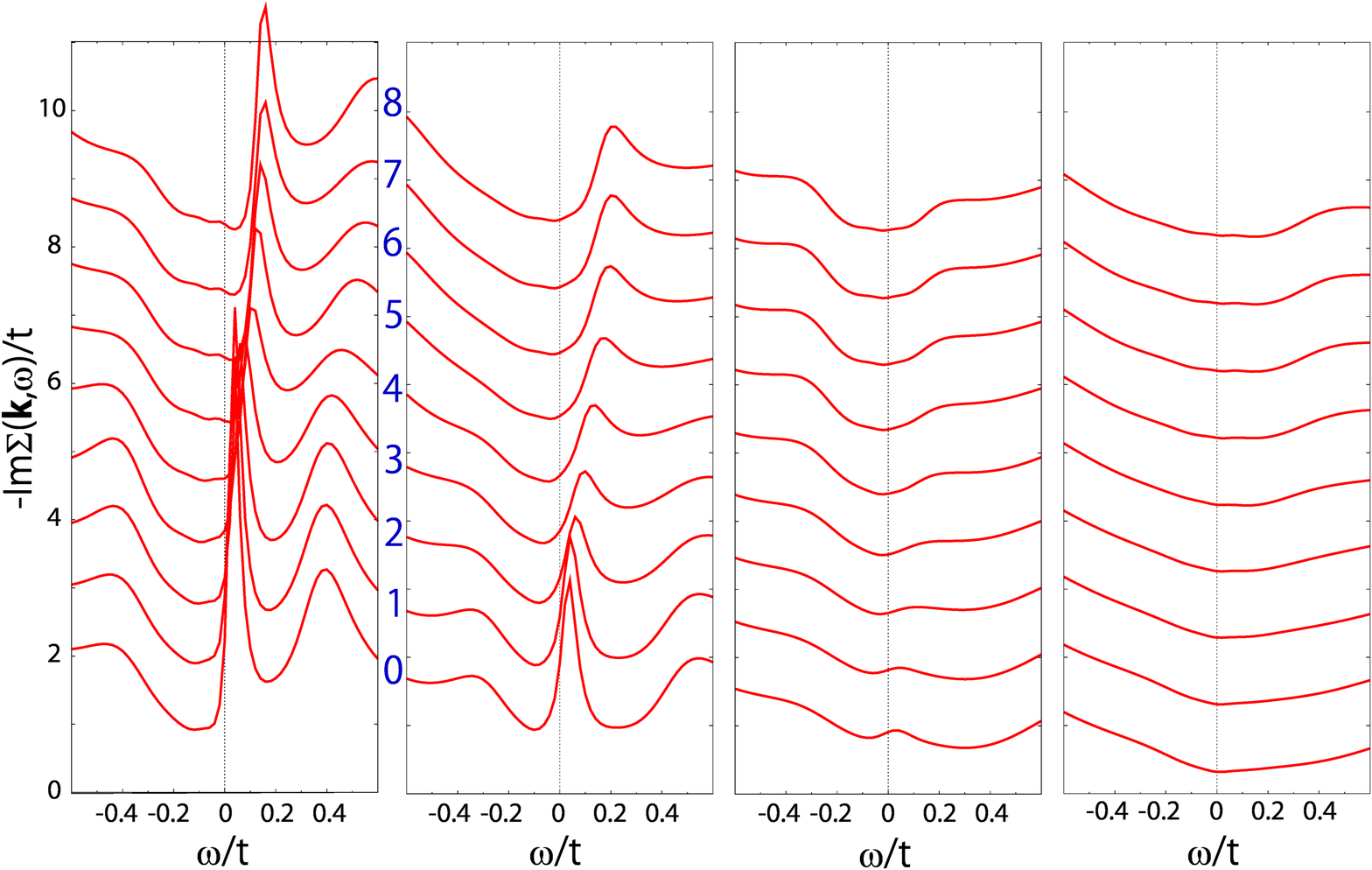}
\label{f2}
\end{center}
\caption{(Color online) The absolute value of the imaginary part of the self-energy calculated at $T=0.04t$ at the same momenta as in Fig. 2. The curves are offset by 1 for clarity.}
\end{figure}

In order to trace the origin of the pseudogap, we display in Fig. 3 the imaginary part of the self-energy at $T=0.04t$ (at the same momenta of Fig. 2).
At a large doping ($n=0.85$) -Im$\Sigma(\Vec{k},\w)$ is a concave function around $\w=0$, with some remnant damping \cite{civelli05,liebsch09,sordi10} which characterizes the width of the quasiparticle peak in $A(\Vec{k},\w)$.
At $n=0.90$ we see a rise of -Im$\Sigma(\Vec{k},\w)$ at an energy slightly above $\w=0$ in the antinodal region.
The peak of -Im$\Sigma(\Vec{k},\w)$ appears to shift to a higher energy toward the nodal region, forming there a shoulder feature.
With further underdoping to $n=0.93$ and $n=0.95$, the peak develops significantly and shows a clearer dispersion.
The large -Im$\Sigma(\Vec{k},\w)$ indicates a strong scattering of electrons and its peak position in the energy-momentum space constitutes a surface of the strong scattering.
In the zero-temperature limit the scattering rate (i.e., -Im$\S$) on this surface is bound to diverge. This gives a surface where the Green's function is zero, which has indeed been found in previous CDMFT studies \cite{stanescu06, stanescu06-2,sakai09,civelli09,sakai10} at zero temperature. In our results the strong peaks of -Im$\Sigma(\Vec{k},\w)$ are the direct consequence of the vicinity to the Mott insulator, whose on-set is typically marked by the appearance of poles in the self-energy. No ordered phase is required to occur in our calculation.

The strong scattering surface is responsible for the formation of the pseudogap in the electronic spectra. Since it disperses to a higher energy while approaching to the nodal point, the one-electron spectra (Fig.2) show a finite gap only above the Fermi level around the node.
The strong-scattering surface is also responsible for a strong renormalization of the band structure (through the real part of the self-energy related to the imaginary part by Kramers-Kronig relations).\cite{stanescu06,stanescu06-2,sakai09,civelli09,sakai10,sakai12,chen12}
While at zero temperature the effect is limited to a relatively narrow energy-momentum region around the surface, at finite temperatures the thermal broadening propagates this effect to a wider range in the energy-momentum space \cite{sakai09-2}.
The quasiparticle peaks are hence pushed to more negative frequenecies at low dopings; as seen in the left panel of Fig.2. 
The intensity of the scattering is about twice stronger in the antinodal region than in the nodal region. The origin of the intensity variation remains to be clarified in a future work. 

\section{Conclusion}

We have explored the doping evolution of the electronic structure of the two-dimensional Hubbard model from the lightly hole-doped pseudogap state to the heavily-doped Fermi liquid state.
The single-electron excitation spectra calculated by the CDMFT show an ``$s$-wave" structure of the pseudogap at low doping, with the gap being effectively shifted above the Ferml level in the nodal region (remaining gapless at the Fermi level).
The pseudogap progressively disappears with doping at all the momenta from the antinode to the node and already vanishes for $n=0.85$ at $T=0.04t$, where well-defined qusiparticle peaks are found in all the directions of the momentum space.
We have identified as the origin of the pseudogap the strong-scattering surface emerging in the vicinity of the Mott insulator.

\end{document}